\documentclass[a4paper,11pt]{article}
\pdfoutput=1 

\usepackage{jheppub} 
\usepackage[compat=1.1.0]{tikz-feynman}
\usepackage{todonotes}
\usepackage{graphicx}
\usepackage{subcaption}
\usepackage[export]{adjustbox}
\usepackage{hyperref}
\usepackage{pifont}
\newcommand{\cmark}{\ding{51}}%
\newcommand{\xmark}{\ding{55}}%

\DeclareRobustCommand{\Sec}[1]{Sec.~\ref{#1}}
\DeclareRobustCommand{\Secs}[2]{Secs.~\ref{#1} and \ref{#2}}

\DeclareRobustCommand{\Fig}[1]{Fig.~\ref{#1}}
\DeclareRobustCommand{\Figs}[2]{Figs.~\ref{#1} and \ref{#2}}

\DeclareRobustCommand{\Eq}[1]{Eq.~(\ref{#1})}
\DeclareRobustCommand{\Eqs}[2]{Eqs.~(\ref{#1}) and (\ref{#2})}

\DeclareRobustCommand{\Ref}[1]{Ref.~\cite{#1}}
\DeclareRobustCommand{\Refs}[1]{Refs.~\cite{#1}}

\newcommand{\ziso}{z_{\rm iso}}

\preprint{MIT-CTP/5011}

\title{Photon isolation and jet substructure}

\author{Zachary Hall}
\author{and Jesse Thaler}

\affiliation{Center for Theoretical Physics, Massachusetts Institute of Technology,\\Cambridge, MA 02139, U.S.A.}

\emailAdd{zachhall@mit.edu}
\emailAdd{jthaler@mit.edu}

\abstract{
We introduce soft drop isolation, a new photon isolation criterion inspired by jet substructure techniques.
Soft drop isolation is collinear safe and is equivalent to Frixione isolation at leading non-trivial order in the small $R$ limit.
However, soft drop isolation has the interesting feature of being democratic, meaning that photons can be treated equivalently to hadrons for initial jet clustering.
Taking advantage of this democratic property, we define an isolated photon subjet:  a photon that is not isolated from its parent jet but is isolated within its parent subjet after soft drop declustering.
The kinematics of this isolated photon subjet can be used to expose the QED splitting function, in which a quark radiates a photon, and we verify this behavior using both a parton shower generator and a perturbative calculation in the collinear limit.
}

\begin{document} 
\maketitle
\flushbottom

\section{Introduction}
\label{sec:1}
Photons produced in high-energy collisions fall into two categories: ``direct'' photons produced in perturbative hard processes and ``indirect'' photons produced from the fragmentation of quark and gluon partons.
Because direct photons access the perturbative part of the collision, they are typically of more interest than indirect photons.
For this reason, photon isolation techniques have been developed to filter out indirect photons, especially from $\pi^0 \rightarrow \gamma \gamma$ decays \cite{
Baer:1990ra, Berger:1990et, Kramer:1991yc, Kunszt:1992np, Glover:1992he,
Buskulic:1995au, Glover:1993xc, GehrmannDeRidder:1997wx,
Frixione:1998jh,
Cieri:2015wwa}.
Although there are different types of isolation criteria used, they all follow roughly the same philosophy: photons collinear to a significant amount of hadronic energy are labeled indirect, while photons well separated from hadronic energy are labeled direct.
By now, photon isolation is a well-established method to study direct photons, with numerous measurements at the Large Hadron Collider (LHC) and previous experiments
\cite{
Glover:1992he,Buskulic:1992ji,Glover:1994he,
Khachatryan:2010fm,Chatrchyan:2011ue,
Aad:2010sp,Aaboud:2017lxm,Aaboud:2017kff%
}.

In the years since the development of photon isolation, jet physics has undergone a rapid evolution, first with the rise of clustering-based jet observables \cite{Catani:1993hr,Ellis:1993tq,Dokshitzer:1997in,Wobisch:1998wt,Wobisch:2000dk,Cacciari:2008gp,Salam:2009jx,Cacciari:2011ma} and more recently with the explosion of the field of jet substructure \cite{Seymour:1991cb,Seymour:1993mx,Butterworth:2002tt,Butterworth:2007ke,Butterworth:2008iy,Abdesselam:2010pt,Altheimer:2012mn,Altheimer:2013yza,Adams:2015hiv,Larkoski:2017jix,Asquith:2018igt}.
Jet substructure provides a rich toolbox to explore soft and collinear dynamics within jets, and it is natural to ask whether substructure techniques could be adapted to handle photons.
At minimum, jet substructure could be used to robustly veto hadronic activity and isolate direct photons.
More ambitiously, jet substructure could facilitate new methods to study indirect photons, by revealing a continuum of collinear photon fragmentation processes from perturbative radiation to hadronic decays.

In this paper, we introduce a new substructure-based photon isolation technique called \emph{soft drop isolation}.
This method derives from soft drop declustering \cite{Larkoski:2014wba}, one of many jet grooming algorithms \cite{Butterworth:2008iy,Ellis:2009su,Ellis:2009me,Krohn:2009th,Dasgupta:2013ihk} that have been successfully adopted at the LHC.
Ordinarily, soft drop declustering is used to identify hard subjets within a jet that satisfy the condition:
\begin{equation}
\label{eq:SDcondition}
    \frac{\text{min}\left(p_{T1},p_{T2}\right)}{p_{T1} + p_{T2}} \geq z_{\text{cut}} \left(\frac{R_{12}}{R_0}\right)^{\beta},
\end{equation}
where $p_{Ti}$ are the transverse momenta of the subjets, $R_{12}$ is their pairwise angular separation, $R_0$ is the jet radius parameter, and $z_{\text{cut}}$ and $\beta$ are the parameters of the soft drop algorithm.
Soft drop isolation \emph{inverts} the condition in \Eq{eq:SDcondition}, thereby selecting ``photon jets'' with no appreciable substructure.
With its origins in jet substructure, soft drop isolation is well suited to the age of particle flow at both CMS \cite{Sirunyan:2017ulk} and ATLAS  \cite{Aaboud:2017aca}.

Like Frixione or ``smooth'' isolation \cite{Frixione:1998jh}, soft drop isolation is collinear safe and fully regulates the collinear divergence of quark-to-photon fragmentation. 
This is in contrast with traditional cone isolation techniques \cite{Baer:1990ra, Berger:1990et, Kramer:1991yc, Kunszt:1992np, Glover:1992he}, which are collinear unsafe.%
\footnote{Traditional cone isolation is collinear unsafe to quark-to-photon fragmentation because of the non-zero energy threshold at zero opening angle.
This is logically distinct from the infrared and/or collinear unsafety of certain cone jet algorithms that make use of unsafe seed axes.}
Collinear-safe photon isolation criteria eliminate the need for parton fragmentation functions \cite{Koller:1978kq, Laermann:1982jr} to regulate the collinear divergence of $q \rightarrow q \gamma$ processes.
This is a significant advantage, as fragmentation functions are inherently non-perturbative and therefore not directly calculable, and experimental measurements \cite{Glover:1993xc, Buskulic:1995au,GehrmannDeRidder:1997wx, Ackerstaff:1997nha, Bourhis:1997yu, Bourhis:2000gs} have significant uncertainties.
For these reasons, collinear-safe photon isolation criteria are preferable for perturbative theoretical calculations.
Note that these statements apply to all orders in perturbative quantum chromodynamics (QCD) but only to leading order in quantum electrodynamics (QED).
Beyond leading order in QED, additional effects such as $\gamma \rightarrow \bar{q} q$ splittings emerge that may require a more delicate treatment (see e.g.~\cite{Frederix:2016ost}).

As we will see, soft drop isolation is equivalent at leading (non-trivial) order to the most common implementation of Frixione isolation, at least when considering the small $R_0$ and small $z_{\text{cut}}$ limits.
Unlike Frixione isolation or cone isolation, though, soft drop isolation is democratic, meaning that it treats photons and hadrons equivalently in the initial clustering step.
This feature is reminiscent of earlier democratic isolation criteria \cite{Buskulic:1995au, Glover:1993xc, GehrmannDeRidder:1997wx}, which can be more natural than undemocratic criteria in cases where jets are the central objects of interest. 
Soft drop isolation is, to our knowledge, the first collinear-safe democratic photon isolation criterion.

\begin{figure}[t]
    \centering
    \begin{subfigure}[b]{0.3\textwidth}
    \centering
    \begin{tikzpicture}
      \begin{feynman}
        \vertex (a) {\(q\)};
        \vertex [right=of a] (b) ;
        \vertex [above right=of b] (f1) {\(\gamma\)};
        \vertex [below right=of b] (c) {\(q\)};
     
        \diagram* {
          (a) -- [fermion] (b) -- [photon] (f1),
          (b) -- [fermion] (c),
        };
      \end{feynman}
    \end{tikzpicture}
    \caption{}
    \label{fig:splitting_a}
    \end{subfigure}
    \hfill
    \begin{subfigure}[b]{0.3\textwidth}
    \centering
    \begin{tikzpicture}
      \begin{feynman}
        \vertex (a) {\(q\)};
        \vertex [right=of a] (b);
        \vertex [above right=1.5cm of b] (f1) {\(\gamma\)};
        \vertex [below right=0.75cm of b] (c);
        \vertex [below right=0.75cm of c] (f3) {\(q\)};
        \vertex [right=0.75cm of c] (f2) {\(g\)}; 
     
        \diagram* {
          (a) -- [fermion] (b),
          (b) -- [photon] (f1),
          (b) -- [fermion] (c),
          (c) -- [gluon] (f2),
          (c) -- [fermion] (f3),
        };
      \end{feynman}
    \end{tikzpicture}
    \caption{}
    \label{fig:splitting_b}
    \end{subfigure}
    \hfill
    \begin{subfigure}[b]{0.3\textwidth}
    \centering
    \begin{tikzpicture}
      \begin{feynman}
        \vertex (a) {\(g\)};
        \vertex [right=of a] (b);
        \vertex [above right=1.5cm of b] (f1) {\(\bar{q}\)};
        \vertex [below right=0.75cm of b] (c);
        \vertex [below right=0.75cm of c] (f3) {\(q\)};
        \vertex [right=0.75cm of c] (f2) {\(\gamma\)}; 
     
        \diagram* {
          (a) -- [gluon] (b),
          (f1) -- [fermion] (b),
          (b) -- [fermion] (c),
          (c) -- [photon] (f2),
          (c) -- [fermion] (f3),
        };
      \end{feynman}
    \end{tikzpicture}
    \caption{}
    \label{fig:splitting_c}
    \end{subfigure}
    \caption{\textbf{(a)} Isolated photon subjet production from a quark at order $\alpha_e$.  The momentum-sharing distribution of this branching in the collinear limit is described by the QED splitting function $P(z)$. \textbf{(b, c)} Processes that contribute to isolated photon subjet production at order $\alpha_e \alpha_s$. Of these, the initial quark term \textbf{(b)} dominates. Not shown are diagrams with a virtual gluon, which are accounted for using the plus prescription.}
    \label{fig:splitting}
\end{figure}

In the second half of this paper, we take advantage of the democratic nature of soft drop isolation to define an isolated photon subjet:  a photon that is not isolated from its parent jet but which is isolated within its parent subjet.
At leading order in the collinear limit, isolated photon subjets arise from the splitting of a quark into a quark plus a photon in QED, as shown in \Fig{fig:splitting_a}.
The probability for a quark to radiate a photon with some angle $\theta_{\gamma}$ and momentum fraction $z_{\gamma}$ is given by:
\begin{equation}
\label{eq:QEDsplit}
    \text{d} P_{q \rightarrow q \gamma} = \frac{\alpha_e e^2}{2 \pi} \,\, \frac{\text{d} \theta_{\gamma}}{\theta_{\gamma}} \,\, P(z_{\gamma}) \,\, \text{d} z_{\gamma}, \qquad 
    P(z) = \left(\frac{1 + (1 - z)^2}{z}\right)_+,
\end{equation}
where $P(z)$ is the (regularized) QED splitting function.
Inspired by related work on the $q \to q g$ splitting function in QCD~\cite{Larkoski:2015lea,Larkoski:2017bvj,Tripathee:2017ybi,Sirunyan:2017bsd,Caffarri:2017bmh,Kauder:2017mhg}, we use isolated photon subjets to expose the QED $q \to q \gamma$ splitting function $P(z)$.
We also investigate the impact of the higher-order $\alpha_s$ corrections in \Figs{fig:splitting_b}{fig:splitting_c}, though we restrict our calculations to the collinear limit.

This work is complementary to earlier experimental investigations of the quark-photon fragmentation function at the Large Electron-Positron Collider (LEP) \cite{Buskulic:1995au, Glover:1993xc, GehrmannDeRidder:1997wx,Ackerstaff:1997nha}.
Notably, \Ref{Ackerstaff:1997nha} exposed the quark-photon fragmentation function down to $z_\gamma \sim 0.2$ by using cluster shape observables to mitigate meson decay backgrounds.
Compared to these studies, the isolated photon subjet approach has the advantage of being perturbatively calculable and likely being easier to implement in the complicated hadronic environment of the LHC.
Additionally, the isolated photon subjet condition regulates higher-order terms such as those in \Figs{fig:splitting_b}{fig:splitting_c}, thereby more directly exposing the QED splitting function as opposed to the inclusive photon fragmentation function.
Similar to the LEP study, the primary background to isolated photon subjets comes from meson decays, but this can be partially controlled using an angular cut on $R_{12}$.

The rest of this paper is organized as follows.
In \Sec{sec:2}, we define soft drop isolation, investigate its features, and analyze its performance in $\gamma$-plus-jet events from a parton shower generator.
In \Sec{sec:3}, we define the isolated photon subjet and compare the extraction of the QED splitting function between a parton shower and an analytic calculation.
We conclude with a discussion of future directions in \Sec{sec:conclusion}.

\section{Photon isolation with soft drop declustering}
\label{sec:2}

Soft drop isolation is based on soft drop declustering, a jet grooming algorithm that removes soft and wide-angle radiation to find hard substructure \cite{Larkoski:2014wba}.
In this section, we show how to tag isolated photons by identifying jets without any substructure.
We first define soft drop photon isolation in \Sec{sec:2.1} and show that it is infrared and collinear safe.
We then show that it is democratic in \Sec{sec:2.2} and compare its behavior to Frixione isolation in \Sec{sec:2.3}.
In \Sec{sec:2.4}, we study soft drop isolation using a parton shower, showing that it performs nearly identically to Frixione isolation.

\subsection{Definition of soft drop isolation}
\label{sec:2.1}

\begin{figure}[t]
    \centering
    \begin{subfigure}[b]{0.47\textwidth}
    \centering
    \begin{tikzpicture}
    		\draw [thick] (0,0) -- (4,0) node[right] {$\gamma$};
     	\draw [thin, dashed] (1,0) -- (1,1) -- (4,1);
    		\draw [thin, dashed] (1.5,0) -- (1.5,-1) --(4,-1);
    		\draw [thin, dashed] (2.5,-1) -- (2.5, -0.75) -- (4, -0.75);
    		\draw [thin, dashed] (2, 0) -- (2,0.5) -- (4,0.5);
    		\draw [thin,dashed] (3.0,0) -- (3.0, -0.25) -- (4.0,-0.25);
    \end{tikzpicture}
    \\[10pt]
    {\Large \color{green}\cmark}
    \caption{}
    \label{fig:sd_pass}
    \end{subfigure}
    \begin{subfigure}[b]{0.47\textwidth}
    \centering
    \begin{tikzpicture}
    		\draw [thick] (0,0) -- (4,0) node[right] {$\gamma$};
     	\draw [thin, dashed] (1,0) -- (1,1) -- (4,1);
    		\draw [thin, dashed] (1.5,0) -- (1.5,-1) --(4,-1);
    		\draw [thin, dashed] (2.5,-1) -- (2.5, -0.75) -- (4, -0.75);
    		\draw [thick] (2, 0) -- (2,0.5) -- (4,0.5);
		\draw [thick] (2.5, 0.5) -- (2.5,0.75) -- (4,0.75);
		\draw [thick] (2.75, 0.5) -- (2.75,0.3) -- (4,0.3);	
    		\draw [thick] (3.0,0) -- (3.0, -0.25) -- (4.0,-0.25);
    \end{tikzpicture}
    \\[10pt]
    {\Large \color{red}\xmark}
    \caption{}
    \label{fig:sd_fail}
    \end{subfigure}
    \caption{Schematic representations of soft drop isolation, where solid (dashed) lines indicate jet constituents kept (dropped) by soft drop declustering. \textbf{(a)}~A photon that \textit{passes} soft drop isolation, because its parent jet fails soft drop, leaving just a singlet photon (as determined by a particle identification scheme). \textbf{(b)}~A photon that \textit{fails} soft drop isolation, because its parent jet has hard substructure that passes soft drop.}
    \label{fig:sd_passfail}
\end{figure}
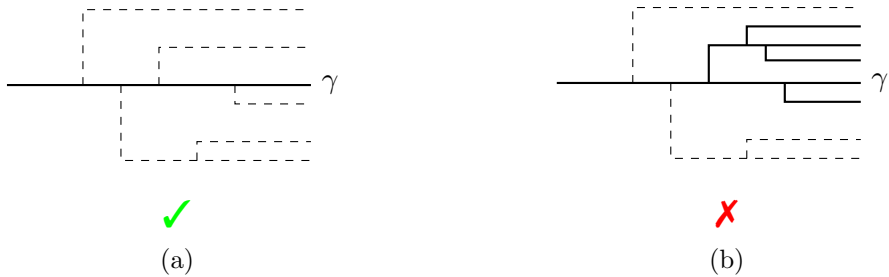

The original soft drop procedure begins with a jet of radius $R$ obtained through some clustering algorithm; this paper uses the anti-$k_t$ algorithm \cite{Cacciari:2008gp} with radius $R = 0.4$ throughout.
Following this, the jet is reclustered using the Cambridge-Aachen (C/A) algorithm \cite{Dokshitzer:1997in,Wobisch:1998wt,Wobisch:2000dk}, yielding an angular-ordered clustering tree.
The jet is then declustered into its two C/A parent subjets; if the soft drop condition in \Eq{eq:SDcondition} is satisfied by the two subjets, then the jet ``passes'' soft drop and is returned as the soft-dropped jet.
Otherwise, the softer (by $p_T$) of the two subjets is dropped and the procedure is repeated on the harder of the two subjets.

As shown in \Fig{fig:sd_passfail}, soft drop isolation is defined in terms of the soft drop algorithm, but with reversed criteria. 
If at no point the jet passes the soft drop condition and one is left with a single constituent that cannot be declustered, then the jet ``fails'' soft drop and the single constituent is returned as the soft-dropped jet.\footnote{Strictly speaking, this corresponds to soft drop in ``grooming mode''~\cite{Larkoski:2014wba}.  In ``tagging mode'', the singlet would simply be vetoed.}
If that single constituent is a photon, then that photon is declared to \textit{pass} soft drop isolation and is labeled as an isolated photon.

Like all photon isolation criteria, soft drop isolation depends on a particle identification scheme to define a (singlet) photon.
This is relevant in the case of prompt photons converted to $e^+\,e^-$ pairs in material, which one typically wants to label as a photon candidate.%
\footnote{When considering electroweak corrections to photon production, it may also be desirable to label vacuum photon-to-lepton splittings $\gamma \rightarrow \ell^+\,\ell^-$ as singlet photons.  We thank Stefano Frixione for discussions on this point.} 
By contrast, one typically wants the particle identification scheme to reject closely collinear $\pi^0 \rightarrow \gamma \gamma$ decays, which can mimic a singlet photon.
In practice, photon definitions are implemented in particle reconstruction algorithms through a combination of cluster-shape observables and tracking \cite{Sirunyan:2017ulk,Aaboud:2017aca}.
For our parton shower study below, we use truth information to label photons, deferring a study of detector effects to future work.

Like soft drop, soft drop isolation depends on the parameters $z_{\text{cut}}$ and $\beta$.
For the algorithm to be collinear safe, we must chose $\beta > 0$.
Although there is some flexibility in choosing these parameters, we will for definiteness use the default parameters:
\begin{equation}
z_{\text{cut}} = 0.1, \qquad \beta = 2.
\end{equation}
Given the matching between the soft drop parameter $z_{\text{cut}}$ and the Frixione parameter $\epsilon$ shown in \Sec{sec:2.3}, these parameter choices are roughly equivalent to the standard ``tight isolation'' parameters outlined in the 2013 Les Houches Accords \cite{Andersen:2014efa}.

We now demonstrate that soft drop isolation is infrared and collinear safe when applied to isolated photons.
The following logic closely follows \Ref{Frixione:1998jh}; a more rigorous proof can be found by following \Refs{Catani:1998yh,Catani:2002ny}.%
\footnote{As in \Ref{Frixione:1998jh}, this isolation criterion would not be safe for simultaneous soft and collinear divergences.  Luckily, this is not relevant for quark and gluon radiation in the presence of a photon, where only one kind of divergence can appear at a time.}
Because soft drop isolation requires the non-photon $p_T$ to vanish as $\Delta R \rightarrow 0$, it is intuitive that collinear divergences will be regulated. 
As seen from \Eq{eq:QEDsplit}, collinear divergences in the process $q \to q \gamma$ have amplitude squared proportional to $1/\theta_\gamma$, where $\theta_\gamma$ is the emission angle.
For a quark with transverse momentum $p_T$ and a photon with transverse momentum $p_{T\gamma}$, the cross section for an isolated photon in the presence of a collinear divergence scales like:
\begin{subequations}
\begin{align}
  \label{eq:collinear_divergence_a}
    \sigma_{\text{SD}}
    & \propto \int \frac{\text{d} \theta_\gamma}{\theta_\gamma} \int \text{d}p^2_T \, \Theta\left[p_{T\gamma}\, \frac{z_{\text{cut}} \left(\frac{\theta_\gamma}{R_0}\right)^{\beta}}{1 - z_{\text{cut}} \left(\frac{\theta_\gamma}{R_0}\right)^{\beta}} - p_T\right], \\[10pt]
    & \sim p_{T\gamma}^2 \,\, \frac{(1 - z_{\text{cut}}) \log (1-z_{\text{cut}})+ z_{\text{cut}}}{\beta (1 - z_{\text{cut}})},
\end{align}
\label{eq:collinear_divergence}%
\end{subequations}
which is clearly convergent.
The Heaviside theta function in \Eq{eq:collinear_divergence_a} is the (inverted) soft drop condition in \Eq{eq:SDcondition}, with the simplifying assumption that $z_{\text{cut}} < \frac{1}{2}$ (which has no effect on the convergence properties).
Just as with Frixione isolation, the fact that soft drop isolation is collinear safe eliminates the dependence of perturbative calculations on fragmentation functions.

Crucially, the soft drop condition does not restrict the phase space of infinitesimally soft gluons, since infinitesimal radiation always satisfies \Eq{eq:SDcondition}.
Infrared divergences from soft gluons have amplitude squared proportional to $1/p_T^2$.
For a gluon with transverse momentum $p_T$, the cross section for an isolated photon in the presence of an infrared divergence scales like:
\begin{subequations}
\begin{align}
    \sigma_{\text{SD}}
    & \propto \int \text{d} \theta_\gamma \int^{p_{T \gamma}^2} \frac{\text{d}p^2_T}{p_T^2} \, \Theta\left[p_{T\gamma}\, \frac{z_{\text{cut}} \left(\frac{\theta_\gamma}{R_0}\right)^{\beta}}{1 - z_{\text{cut}} \left(\frac{\theta_\gamma}{R_0}\right)^{\beta}} - p_T\right],
    \label{eq:infrared_divergence_1}%
    \\[10pt]
    & \sim R_0\,\left(\log\left(z_{\text{cut}}\right) - \beta\right),
\label{eq:infrared_divergence_2}%
\end{align}
\label{eq:infrared_divergence}%
\end{subequations}
which is again convergent.
In \Eq{eq:infrared_divergence_2}, we have used the plus prescription to perform the integral over $p_T$, which is valid since we have not restricted the phase space of infinitesimally soft gluons and thereby ensured that real-virtual cancellation will occur.

Because soft drop isolation is based on declustering, it is easy to check that infrared and collinear safety persists with multiple emissions.
Each step in the declustering procedure acts on two subjets, so the way the algorithm handles divergence structures will be the same at each step.
In this way, soft drop isolation gives an infrared- and collinear-safe definition for isolated photons.

\subsection{Soft drop isolation is democratic}
\label{sec:2.2}

The terms ``democratic isolation'' and ``the democratic approach'' have typically referred to a particular form of isolation pioneered in the LEP era for the study of the photon fragmentation function \cite{Buskulic:1995au, Glover:1993xc, GehrmannDeRidder:1997wx}.
In traditional democratic isolation, the entire event is clustered into jets, including both photons and hadrons. 
This step, which treats photons and hadrons equally, is the origin of the term ``democratic''; undemocratic criteria such as Frixione isolation and cone isolation instead center the isolation scheme around the photon.
Following the jet clustering step, a photon is defined to be isolated if it accounts for the majority of the energy of its parent jet.
However, traditional democratic isolation is essentially just a clustering-based form of cone isolation and correspondingly suffers from the same problem of collinear unsafety.

As is clear from the definition in \Sec{sec:2.1}, soft drop isolation is a democratic criterion.
Much like traditional democratic isolation, soft drop isolation begins by clustering the particles in an event democratically into jets.
It is only after the jet has been completely declustered that the soft drop isolation algorithm distinguishes between photons and other particles.
Unlike traditional democratic isolation, though, soft drop isolation is collinear safe.
We believe that soft drop isolation is the first democratic collinear-safe photon isolation criterion.

As a democratic criterion, the logic of soft drop isolation is different from that of undemocratic criteria.
Instead of testing whether a photon is isolated, soft drop isolation tests whether a jet contains an isolated photon.
Democratic isolation techniques are thus more natural for cases where one is testing for multiple isolated photons or for cases where jets are the most natural object.
Frixione isolation or cone isolation, on the other hand, are more natural for testing the hardest photon in an event to see if it is isolated. 

The fact that soft drop isolation is democratic leads to some mild differences with Frixione isolation.
The reasons for this are twofold.
First, the fact that the photon is isolated from a jet with radius $R$ means that this isolation radius is not strictly drawn around the photon: the photon might not be exactly at the jet center.
Therefore, there can be some differences when the photon is off-center and there are hard features at a distance $\sim R$ from the photon.
This has little effect in practice, however, since isolated photons naturally contain most of the momentum of the jet and therefore appear very close to the jet center.
Second, soft drop isolation is applied after the event has already been clustered into jet objects, whereas Frixione isolation is applied before the event has been clustered.
Frixione isolation thus can allow low-momentum objects at angles $\Delta R < R_0$, whereas such objects are mostly excluded by soft drop isolation (namely, they can only occur due to deviations of the photon from the jet center).
These differences between democratic and undemocratic approaches will be explored further in \Sec{sec:2.4}.

Soft drop's democratic nature makes it a natural choice for the study of jet structure and substructure.
The isolated photon subjet introduced later in \Sec{sec:3.1} is one such example that would be quite unnatural to define with a non-democratic criterion.
More broadly, democratic criteria are the natural choice for modern hadron colliders, where jets are ubiquitous objects and clustering techniques like anti-$k_t$  \cite{Cacciari:2008gp} are now used by default.
\subsection{Relationship to Frixione isolation}
\label{sec:2.3}

Given the above discussion, it is perhaps surprising that (democratic) soft drop isolation turns out to be equivalent to (undemocratic) Frixione isolation, at least in a particular limit.
For small $R_0$ and small $z_{\text{cut}}$, there are appropriate choices of soft drop parameters such that soft drop isolation and the most common form of Frixione isolation impose the same restriction on two-particle final states.
Since this corresponds to the leading (non-trivial) order configuration in \Fig{fig:splitting_a}, we say that the two criteria are equivalent at leading order.

Frixione or ``smooth'' isolation \cite{Frixione:1998jh} has been the preferred photon isolation criterion for perturbative calculations.
In contrast to cone isolation, Frixione isolation regulates the collinear divergence by forcing the partonic energy to zero in the collinear limit.
In this way, the exact collinear divergence from $q \to q \gamma$ is fully eliminated without in any way restricting the soft phase space, which is required in order to ensure real and virtual cancellation of soft gluon divergences.

Frixione isolation uses an initial angular cut at some radius from the photon $R_0$.
The particles within that radius are then required to pass a momentum cut based on an angular function $X(\Delta R)$, typically called a Frixione function.
The full condition may be expressed in terms of the transverse momentum $p_{T i}$ and distance to the photon $R_{i,\gamma}$ of each hadronic particle as:%
\footnote{Implementations of Frixione isolation often use the transverse energy $E_T$ in place of the transverse momentum $p_T$.  Given the ambiguities in defining transverse energy and the assumption of high energies, we will instead use $p_T$ throughout.}
\begin{equation}
    \forall\, \Delta R \leq R_0:\quad \sum_{i} \,\, p_{T i} \,\, \Theta\left(\Delta R - R_{i,\gamma}\right) < \, X(\Delta R).
\end{equation}
There is significant flexibility in the choice of Frixione function $X(\Delta R)$.
The most common function used in the literature \cite{Frixione:1998jh,Cieri:2015wwa,Binoth:2010nha,AlcarazMaestre:2012vp,Andersen:2014efa,Aaboud:2017lxm,Aaboud:2017kff} is:
\begin{equation}
\label{eq:frix_x(R)}
X(\Delta R) = p_{T \gamma} \, \epsilon \, \left(\frac{1 - \cos(\Delta R)}{1 - \cos(R_0)}\right)^n.
\end{equation}
Under the ``tight isolation'' parameters outlined in the Les Houches Accords \cite{Andersen:2014efa}, typical parameter values are $\epsilon \sim 0.1$ and $n = 1$. Another common implementation \cite{Cieri:2015wwa, Andersen:2014efa} uses a fixed $E_T^{\rm iso}$ in place of $p_{T \gamma} \, \epsilon$ in \Eq{eq:frix_x(R)}.

At leading order (corresponding to one additional particle within the photon's isolation cone and taking the small $R_0$ limit), the Frixione isolation condition in \Eq{eq:frix_x(R)} becomes:
\begin{equation}
\label{eq:frixcosleading}
    p_T < p_{T\gamma}\, \epsilon \left(\frac{\Delta R}{R_0}\right)^{2 n}.
\end{equation}
It should be noted that this form of $X(\Delta R)$ is equivalent to another Frixione function described in \Ref{Frixione:1998jh}, though this function has not found widespread implementation.

Looking at \Eq{eq:collinear_divergence_a}, the leading-order soft drop criterion with $z_{\text{cut}} < \frac{1}{2}$ is:
\begin{equation}
\label{eq:sdleading}
    p_T < p_{T \gamma} \frac{z_{\text{cut}} \left(\frac{\Delta R}{R_0}\right)^{\beta}}{1 - z_{\text{cut}} \left(\frac{\Delta R}{R_0}\right)^{\beta}}.
\end{equation}
This is clearly equivalent to \Eq{eq:frixcosleading} in the small $z_{\text{cut}}$ or $\frac{\Delta R}{R_0}$ limits with the identification $z_{\text{cut}} = \epsilon$ and $\beta = 2 n$.
We should also note that, given the flexibility in choosing a Frixione function, it is possible to choose $X(\Delta R)$ corresponding exactly to the right-hand side of \Eq{eq:sdleading}.
This form of Frixione isolation would be fully equivalent to soft drop isolation at leading order.%
\footnote{This equivalence gives another way to understand why, with appropriate choice in parameters, soft drop isolation is safe to infrared and collinear divergences.
Just like Frixione isolation, soft drop isolation fully eliminates collinear fragmentation without restricting the soft gluon phase space.
}

Despite the leading-order equivalence of Frixione and soft drop isolation, there are important differences at higher orders.
These differences stem from the fact that soft drop isolation is based on clustering, whereas Frixione isolation is based on a more traditional cone approach.
The details of which scheme is stricter depend on the precise phase space configuration, and it is not possible to make a general statement about the differences in multi-particle configurations.

In practice, differences due to higher-order configurations are negligible in most realistic settings, as seen in the parton shower study below.
Additionally, we found that the two schemes closely matched even with the differences between \Eqs{eq:frix_x(R)}{eq:sdleading} at $\Delta R \sim R_0$.
Instead, the primary differences between the two schemes stem from the fact that soft drop isolation is democratic, as already discussed in \Sec{sec:2.2}.

\subsection{Parton shower study}
\label{sec:2.4}
 
As a practical test of soft drop isolation, we now perform a parton shower study of isolated photon production in the $\gamma$+jet(s) final state.
Not surprisingly given their leading-order equivalence, we find that soft drop and Frixione isolation perform nearly identically, though soft drop isolation's democratic construction leads to some differences in angular distributions.

We generated events in \textsc{Pythia} 8.223 \cite{Sjostrand:2006za,Sjostrand:2014zea} from proton-proton collisions with center-of-mass energy 13 TeV, using the default settings for hadronization and underlying event.
We created a sample of 800,000 events from the \textsc{Pythia} \texttt{PromptPhoton} process, which encodes Compton-like processes that produce a hard photon.%
\footnote{To ensure that there were sufficient events at high photon ${p}_T$, we used binned event generation with bin edges imposed on the hard process of $\hat{p}_T = (100,200,300,400,600,800,1000,1500,\infty)$ GeV.  The events were then reweighted proportional to the generated cross sections.}
In total, \textsc{Pythia} produces photons from the hard scattering process, initial-state radiation (ISR), final-state radiation (FSR), and final-state hadron decays (primarily from neutral pions).
Though not shown, we also tested a similar sample of \texttt{HardQCD} events, which encodes $2 \rightarrow 2$ QCD processes that can produce isolated photons from extra initial-state or final-state emissions; the results did not offer any new qualitative insights compared to the \texttt{PromptPhoton} sample.
Jet clustering and photon isolation were performed using \textsc{FastJet} 3.2.1 \cite{Cacciari:2011ma}.
Soft drop was implemented using the \textsc{FastJet Contrib} 1.026 \texttt{RecursiveTools} package \cite{fastjetcontrib}.

For our event selection, we require an isolated photon with $p_{T\gamma} > 125$ GeV and one hadronic jet with $p_{T\text{jet}} > 100$ GeV.
We use the condition $p_{T X} > 25$ GeV to define any additional jets that might appear in the event.
A rapidity cut of $|y| < 2$ was applied to the final photon and jet objects after jet clustering.
These selection criteria were chosen to roughly match a photon isolation study from ATLAS \cite{Aaboud:2017kff}.
For each isolation criterion, we use the tight isolation parameters: $z_{\text{cut}} = \epsilon = 0.1$, $\beta = 2n = 2$, and $R_0 = 0.4$ \cite{Andersen:2014efa}. 
We used \textsc{Pythia} truth information to perform particle identification.

Because of the democratic versus undemocratic distinction, we had to use slightly different photon selection schemes for soft drop and Frixione isolation.
For soft drop isolation, we first clustered the event into $R = 0.4$ jets with $p_{T X} > 25$ GeV and tested each jet for an isolated photon with $p_{T\gamma} > 125$ GeV and $|y_{\gamma}| < 2$; the remaining hadrons from the isolated-photon jet were discarded.
For Frixione isolation, every photon with $p_{T\gamma} > 125$ GeV and $|y_{\gamma}| < 2$ was tested for isolation; if such a photon was found, then the rest of the event was clustered into $R = 0.4$ jets.
In the case where an event contained multiple isolated photons, we used only the hardest isolated photon.

\begin{figure}[t]
    \centering
    \begin{subfigure}[b]{0.47\textwidth}
    \centering
    \includegraphics{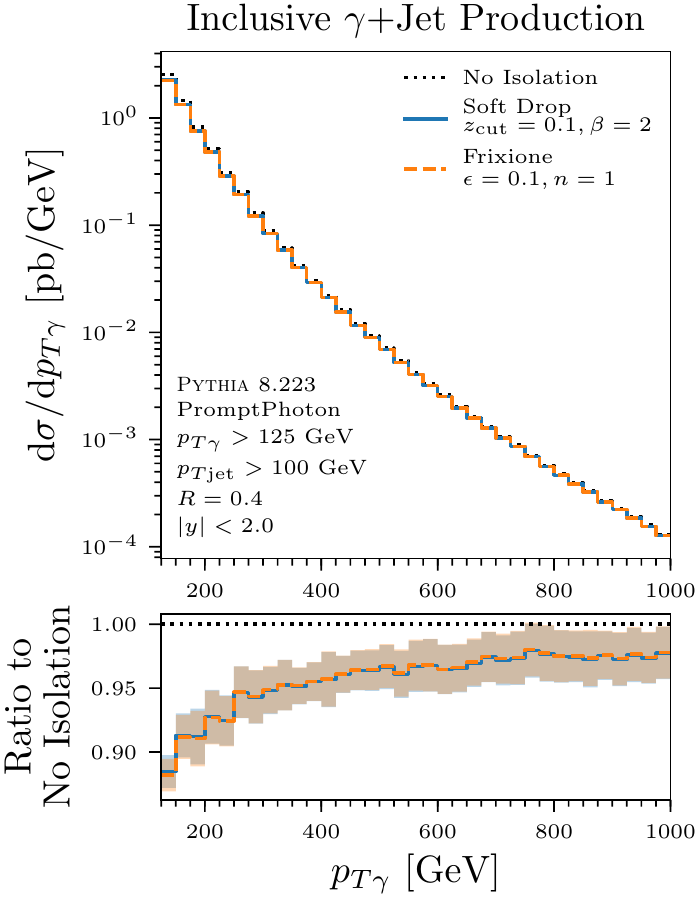}
    \caption{}
    \label{fig:promptpt}
    \end{subfigure}
    \hfill
    \begin{subfigure}[b]{0.47\textwidth}
    \includegraphics{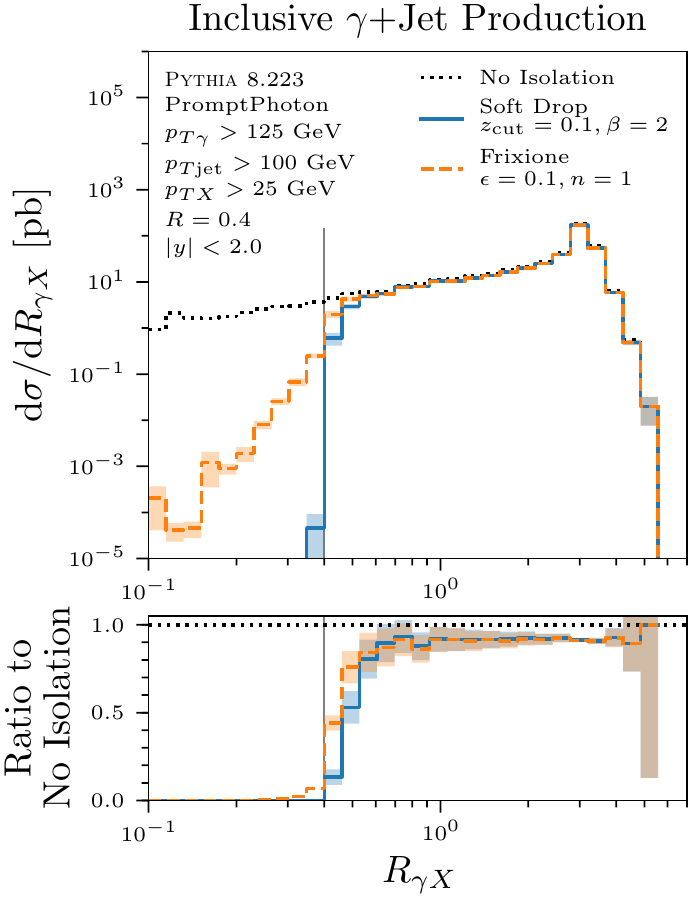}
    \caption{}
    \label{fig:promptdr}
    \end{subfigure}
    \caption{
    Inclusive $\gamma$+jet production cross sections from the \textsc{Pythia} \texttt{PromptPhoton} process, comparing the spectrum of soft drop isolation, Frixione isolation, and the hardest photon without isolation.
    \textbf{(a)} Photon transverse momentum $p_{T \gamma}$.    
    \textbf{(b)} Angle $R_{\gamma X}$ between the photon and the nearest object with $p_{T X} > 25$ GeV.
    In both figures, the bottom panels show the ratios to the non-isolated case, and the shading indicates statistical uncertainties. 
    Although the $p_{T \gamma}$ spectra are nearly identical, there are significant differences in the $R_{\gamma X}$ spectra due to soft drop isolation's democratic nature.}
\end{figure}

In \Fig{fig:promptpt}, we show the photon $p_T$ spectrum for each isolation scheme, as well as for the hardest photon (isolated or not) in each event.
The soft drop and Frixione distributions are nearly identical, showing that the differences between soft drop and Frixione isolation arising from higher-order effects mentioned in \Sec{sec:2.3} are extremely small in practice.
There are on average ~5\% differences between the isolated photon spectra and the hardest photon spectrum, indicating that both isolation schemes properly identify direct photons.
Notably, the two isolated spectra exhibit average differences of less than 0.1\% (below the precision of this study), showing that the soft drop isolation and Frixione isolation perform nearly identically.

In \Fig{fig:promptdr}, we show the angular distance $R_{\gamma X}$ between the isolated photon and the nearest inclusive jet with $p_{T X} > 25$ GeV and $|y_X| < 2$.
As expected, the isolated photon spectra are significantly reduced compared to the non-isolated spectrum for $R_{\gamma X} < 0.4$.
The soft drop and Frixione distributions are very similar for $R_{\gamma X}$  much larger than $0.4$, but there are significant differences between the two isolation schemes in the transition region around $R_{\gamma X} = 0.4$.

For $R_{\gamma X} < 0.4$, these differences are not due to any differences in strictness but rather to soft drop's democratic construction.
Because in Frixione isolation the clustering happens after the isolation step, it is possible for low-energy objects within the photon's isolation cone to become part of one of the inclusive jets $X$.
In contrast, soft drop isolation performs the clustering before the isolation step.
Therefore, the only cases in which $R_{\gamma X} < 0.4$ would be permitted are those where the photon is significantly off-center from the jet axis.
These cases are exceedingly rare, and as such, the soft drop isolation spectrum exhibits a relatively hard cutoff at $R_{\gamma X} = 0.4$.
We suspect that this hard cutoff behavior will be desirable for future direct photon studies at the LHC.

For $R_{\gamma X} \simeq 0.4$, soft drop isolation is more strict than Frixione isolation due to the difference in defining an isolation region through clustering versus through cones.
In soft drop isolation, hard objects at $R_{\gamma X}$ slightly greater than 0.4 will often cluster with the photon.
In Frixione isolation, by contrast, hard objects at this distance will not factor into the isolation, as they fall outside of the isolation cone.
The result is that we expect soft drop isolation to be somewhat stricter in such configurations.
This can be observed in \Fig{fig:promptdr}, where the soft drop isolation spectrum is suppressed relative to Frixione isolation in the approximate region $0.4 < R_{\gamma X} < 0.7$.

We used \textsc{Pythia} truth information to analyze the performance of each isolation scheme as applied in the above study.
Although in the plots above we used only the hardest isolated photon in the event, the following efficiency values include all photons that passed the initial $p_T$ and $y$ cuts.
Soft drop isolation and Frixione isolation each had around 90\% efficiency of tagging direct photons as prompt photons.
Both isolation criteria achieved 100\% rejection of indirect photon backgrounds from final-state hadron decays (limited by the statistics of our sample).
For FSR, which can generate photons both collinear to and well separated from jets, we analyzed both wide-angle radiation, defined as emissions with angle $\Delta R > 0.4$, and collinear radiation, defined as emissions with angle $\Delta R < 0.4$.
Both isolation criteria tagged 53\% of photons from wide-angle FSR as prompt and achieved more than 99\% rejection of collinear FSR.

The above study validates the use of soft drop isolation to identify direct photons.
In the context of \textsc{Pythia}, the level of background rejection from both isolation criteria is so high that it was difficult to get a trustable sample of isolated photons from collinear FSR or hadron decays.
Although the above analysis indicates that soft drop isolation and Frixione isolation give very similar indirect photon background rates when using the tight isolation parameters, a detailed study with a detector simulation (including particle identification that accounts for photon conversion and collinear pion decays) would be needed to fully quantify the differences.

\section{Exposing the QED splitting function}
\label{sec:3}

Because soft drop isolation is democratic, we can naturally use it in contexts where photons play a key role in the substructure of a jet.
The goal of this study is to use the kinematics of isolated photon subjets to expose the QED $q \to q \gamma$ splitting function.
We first give a concrete definition of an isolated photon subjet in \Sec{sec:3.1}.
We then calculate the kinematics of the isolated photon subjet to order $\alpha_e$ in the collinear limit in \Sec{sec:3.2} and show that the photon momentum fraction is directly given by the QED splitting function.
We extend this calculation to order $\alpha_e \alpha_s$ in \Sec{sec:3.3} and show that the qualitative features do not change.
In \Sec{sec:3.4}, we test this procedure with a parton shower generator, where we find behavior consistent with the analytic calculations.

\subsection{Definition of an isolated photon subjet}
\label{sec:3.1}

Our definition of an isolated photon subjet uses a combination of soft drop declustering and soft drop isolation to identity a quark-like jet with photon substructure.
We begin with a jet of radius $R$ obtained through some clustering algorithm (anti-$k_T$ in our study). 
Soft drop is then applied to the jet with $z_{\text{cut}} = 0.1$, $\beta = 0$, and radius parameter $R_0 = R$, such that soft drop acts like the modified Mass Drop Tagger (mMDT) \cite{Dasgupta:2013ihk}.
Events that pass this step now have two prong substructure, and analogous to the QCD splitting function study of \Refs{Larkoski:2015lea,Larkoski:2017bvj}, the choice $\beta = 0$ ensures that the $z$ distribution of the resulting subjets is not biased. 
We then decluster the soft-dropped jet into its two constituent subjets and apply soft drop isolation to each subjet with $z_{\text{cut}} = 0.1$, $\beta = 2$, and radius parameter $R_0 = R_{12}/2$.%
\footnote{We also performed a study using $R_0 = R$ in the soft drop isolation criterion (while still applying the isolation only to the subjet constituents); although this version of the criterion does lead to sensible results, we found it to be more sensitive to non-perturbative hadronization effects.}
If exactly one of the subjets passes soft drop isolation, it is labeled as an isolated photon subjet.

\begin{figure}[t]
	\hspace{3.2cm}
    \includegraphics[left]{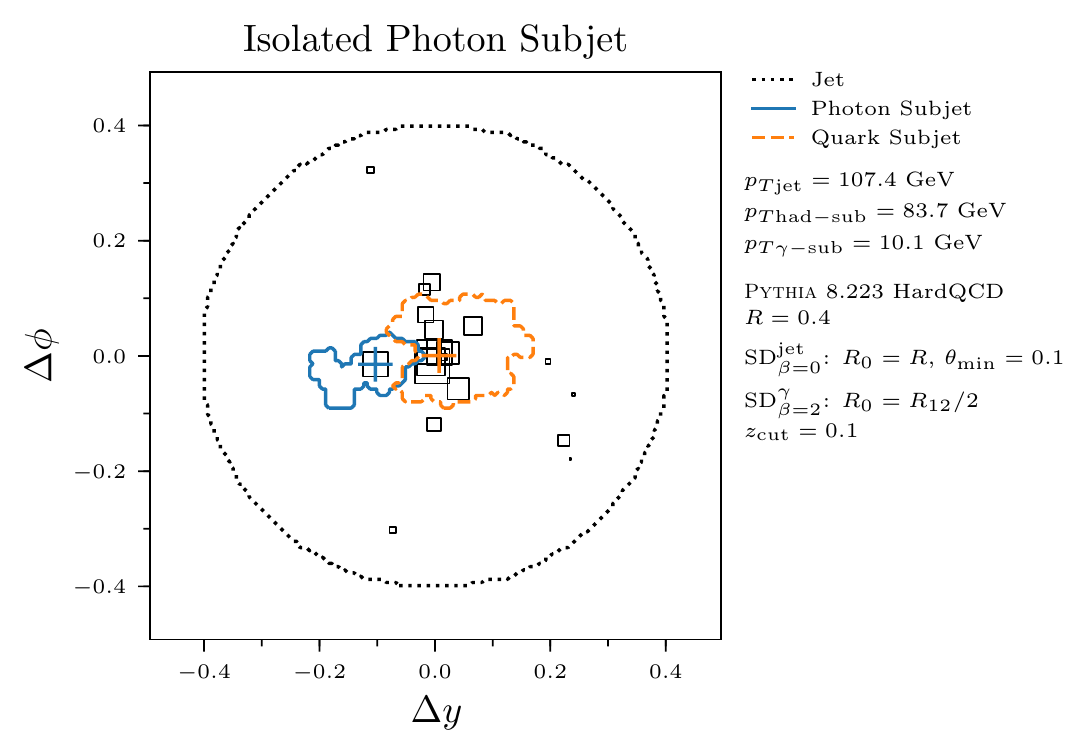}
    \caption{Example jet with an isolated photon subjet from a $q \rightarrow q \gamma$ spltting. For the initial soft drop, denoted $\rm{SD}^{\rm jet}_{\beta = 0}$ (equivalent to mMDT \cite{Dasgupta:2013ihk}), we used parameters $z_{\rm cut} = 0.1$, $\beta = 0$, and $R_0 = R = 0.4$. For the subjet isolation criterion, denoted $\rm{SD}^{\gamma}_{\beta = 2}$, we used parameters $z_{\rm cut} = 0.1$, $\beta = 2$, and $R_0 = R_{12}/2$, where $R_{12}$ is the angle between the two subjets.}
    \label{fig:jetimage}
\end{figure}

In \Fig{fig:jetimage}, we show an example jet from \textsc{Pythia} that contains an isolated photon subjet.
The details of the event generation will be given in \Sec{sec:3.4}.
We see that the first step of soft drop declustering has decreased the active area \cite{Cacciari:2008gn} from the original jet (black, dotted) to the blue and orange subjets.
The blue subjet consists of only a single photon.
The orange, dashed subjet arises from the showering and hadronization of a quark parton.
Using the \textsc{Pythia} event record, we can verify that this configuration does indeed arise from a $q \rightarrow q \gamma$ splitting.

The momentum fraction of the isolated photon subjet provides a novel way to expose the QED splitting function, both in perturbative calculations and in experiment. 
The QED splitting function, given in \Eq{eq:QEDsplit}, describes the probability distribution of the momentum sharing $z$ between the photon and the quark.
We define the isolated photon momentum sharing as
\begin{equation}
\label{eqn:def-ziso}
\ziso = \frac{p_{T \text{$\gamma$-sub}}}{p_{T\text{$\gamma$-sub}} + p_{T \text{had-sub}}},
\end{equation}
as a proxy for the partonic $z$, where $p_{T \text{$\gamma$-sub}}$ is the transverse momentum of the isolated photon subjet and $p_{T \text{had-sub}}$ is the transverse momentum of the other (hadronic) subjet.%
\footnote{
	We also performed a study using $p_{T \gamma}/p_{T\text{jet,SD}}$ as the proxy for partonic $z$, where $p_{T \gamma}$ is the transverse momentum of the photon as opposed to the entire isolated photon subjet and $p_{T\text{jet,SD}}$ is the transverse momentum of the soft-dropped jet. We elected to use the definition in \Eq{eqn:def-ziso} because it ensures a hard cutoff at $\ziso = z_{\rm cut}$ and because it is less sensitive to the effects of hadronizaton.
}
In order to eliminate the primary background from meson decays, we implemented a simple cut on the angle between the two subjets $R_{12} > \theta_{\text{min}}$; a similar cut was used in the CMS study of the QCD splitting function~\cite{Sirunyan:2017bsd}.
The details of this cut are discussed further in \Sec{sec:3.4}.
Note that with this $\theta_{\text{min}}$ restriction, the $\ziso$ observable is infrared and colllinear safe, not just Sudakov safe~\cite{Larkoski:2013paa,Larkoski:2015lea}.

\subsection{Order $\alpha_e$ calculation}
\label{sec:3.2}

We now calculate the differential cross section in $\ziso$ to lowest non-trivial order, focusing on the collinear limit in the fixed-coupling approximation.
At order $\alpha_e$, the cross section is quite simple to evaluate.
There is only one term that contributes, corresponding to the single quark-photon branching from \Fig{fig:splitting_a}.
The cross section can be expressed in terms of the initial quark cross section $\sigma_q$, the quark charge $e_q$, the emission angle $\theta_{\gamma}$, the momentum sharing $z_{\gamma}$, and the order $\alpha_e$ isolated photon subjet condition $\Theta_{(1,0)} $ as:
\begin{equation}
\label{eq:lowestordercrosssection}
    \frac{\text{d}\sigma_{(1,0)}}{\text{d} \ziso} = \,
    \int \text{d}\sigma_q \,\, \frac{\alpha_e e_q^2}{2 \pi} \,\,\frac{\text{d} \theta_{\gamma}}{\theta_{\gamma}} \,\, \text{d} z_{\gamma} \, P(z_{\gamma})\,\,\Theta_{(1,0)} ,
\end{equation}
where the notation $(m,n)$ refers to the order $\alpha_e^m \alpha_s^n$.

Because at this order the jet consists of only a quark and a photon, the procedure in \Sec{sec:3.1} always identifies a quark subjet and a photon subjet, which is automatically an isolated subjet.
The only conditions are that the two particles fall within the jet radius, that the jet as a whole pass the initial soft drop condition, and that the two subjets pass the minimum relative-angle condition:
\begin{equation}
    \Theta_{(1,0)} = 
    \Theta\left[z_{\gamma} - z_{\text{cut}}\right]
    \Theta\left[\left(1 - z_{\gamma}\right) - z_{\text{cut}}\right]
    \delta\left[\ziso - z_{\gamma}\right]
    \Theta\left[R - \theta_{\gamma}\right]
    \Theta\left[\theta_{\gamma} - \theta_{\text{min}}\right].
\end{equation}
Inserting this into \Eq{eq:lowestordercrosssection}, our cross section neatly factorizes into angular and momentum-fraction components, yielding a $\ziso$ distribution that is directly proportional to the splitting function:
\begin{equation}
\begin{split}
\label{eq:lowestorderfinalanswer}
    \frac{\text{d} \sigma_{(1,0)}}{\text{d} \ziso} &=
    \sigma_q \frac{\alpha_e e_q^2}{2 \pi} \, \int_{\theta_{\text{min}}}^{R} \, \frac{\text{d} \theta_{\gamma}}{\theta_{\gamma}} \, \int_{z_{\text{cut}}}^{1 - z_{\text{cut}}}\, \text{d} z_{\gamma} \, P(z_{\gamma}) \,
        \delta\left[\ziso - z_{\gamma}\right]\\
    &= \sigma_q \frac{\alpha_e e_q^2}{2 \pi} \, \log\left(\frac{R}{\theta_{\text{min}}}\right) \,\,P(\ziso) \,\Theta\left[\ziso - z_{\text{cut}}\right]\,\Theta\left[1 - z_{\text{cut}} - \ziso \right].
\end{split}
\end{equation}
Thus, at order $\alpha_e$ the isolated photon subjet momentum fraction directly exposes the QED $q \to q \gamma$ splitting function.

The initial quark cross section $\sigma_q$ is the cross section for quark jet production at the $p_T$ scale of the calculation. At order $\alpha_e$, $\sigma_q$ appears only as a factor in normalization; at order $\alpha_e \alpha_s$, where both quark jet and gluon jet terms contribute, the ratio of $\sigma_q$ to its gluon jet production counterpart $\sigma_g$ is relevant. These values are discussed in detail in \Sec{sec:3.3}.

\subsection{Order $\alpha_e \alpha_s$ calculation}
\label{sec:3.3}

Going to higher orders, one might worry that the simple behavior in \Eq{eq:lowestorderfinalanswer} would be spoiled by QCD radiation within the jet.
This turns out not to be the case.
The reason is that the isolated photon subjet condition regulates singularities collinear to the photon, such that higher-order terms in the inclusive parton-photon fragmentation function are controlled without diminishing the order $\alpha_e$ splitting function.
Although there are still higher-order corrections, they are significantly reduced compared to the raw fragmentation function. 
In this way, the isolated photon subjet more directly exposes the QED splitting function instead of merely exposing the parton-photon fragmentation function.

We can verify the above statements by performing a calculation of the $\ziso$ distribution at order $\alpha_e \alpha_s$.
At this order, analytic calculations of the cross section become considerably more involved, even restricting to the collinear limit with fixed coupling and strongly-ordered emissions.
Two terms contribute to the cross section: the case in which an initial quark emits a photon and a gluon (\Fig{fig:splitting_b}), and the case in which an initial gluon splits into a quark-antiquark pair, one of which then radiates a photon (\Fig{fig:splitting_c}).
Of these two terms, the initial-quark case is dominant, as the initial gluon will be almost entirely excluded by the subjet isolation step.

We work in the strongly-ordered limit, with the emission ordering determined by a generalized virtuality $Q = z(1 - z)\theta^n$.
By changing the value of $n$, we can get a sense of the uncertainties in our calculation, though we emphasize that we have not performed a comprehensive uncertainty estimate.
The choice $n = 1$ corresponds to $k_t$ ordering, $n = 2$ corresponds to a mass ordering, and we also test $n = 1/2$ for completeness.
For the initial-quark diagram in \Fig{fig:splitting_b}, the ordering determines whether the gluon or the photon is emitted first.
For the initial-gluon diagram in \Fig{fig:splitting_c}, the gluon-to-quarks splitting is required to occur first.

The total differential cross section in the observable $\ziso$ can be expressed in terms of the initial-quark cross section $\sigma_q$, the initial-gluon cross section $\sigma_g$, each emission's angle $\theta$ and momentum sharing $z$, the azimuthal angle with respect to the jet axis between emissions $\phi$, the $q \rightarrow q \gamma$ and $q \rightarrow q g$ splitting function $P$, the $g \rightarrow q \bar{q}$ splitting function $P_{qg}$, and the order $\alpha_e \alpha_s$ isolated photon subjet condition $\Theta_{(1,1)}$:\footnote{The name $z_g$ for the momentum fraction of the gluon should not be confused with the groomed momentum fraction from \Ref{Larkoski:2015lea}.}
\begin{equation}
\begin{split}
    \frac{\text{d}\sigma_{(1,1)}}{\text{d} \ziso} = \,&\int
    \text{d}\sigma_q \,\, \frac{\alpha_e e_q^2}{2 \pi} \,\frac{\text{d} \theta_{\gamma}}{\theta_{\gamma}} \, \text{d} z_{\gamma} \, P(z_{\gamma})
    \,\,
    \frac{\alpha_s C_F}{2 \pi} \,\frac{\text{d} \theta_{g}}{\theta_{g}} \,\text{d} z_{g} \, P(z_{g}) 
    \,\,
    \frac{\text{d} \phi}{2\pi} \,\Theta_{(1,1)} \left[p_q, p_g, p_{\gamma}\right] \\[10pt]
    ~+2\,&\int
    \text{d}\sigma_g \,\, 
    \frac{\alpha_s T_F}{2 \pi} \,\frac{\text{d} \theta_{q}}{\theta_{q}} \,\text{d} z_{q} \, P_{qg}(z_{q})
    \,\,
    \frac{\alpha_e e_q^2}{2 \pi} \,\frac{\text{d} \theta_{\gamma}}{\theta_{\gamma}} \, \text{d} z_{\gamma} \, P(z_{\gamma})
    \,\,
    \frac{\text{d} \phi}{2\pi} \,\Theta_{(1,1)} \left[p_q, p_{\bar{q}}, p_{\gamma}\right].
\end{split}
\label{eq:aeascross}
\end{equation}

For simplicity of presentation, we do not give the precise functional form for $\Theta_{(1,1)}$.
This function contains the clustering, initial soft drop, and subjet isolation steps and depends on the four-momenta of the final-state particles.
These four-momenta in turn depend on how the branching variables are mapped to physical kinematics.
We decide to define four-momenta by conserving three-momentum at each branching; we do not conserve energy in this process, which is consistent in the collinear limit.
For the branching $A \to BC$ of a particle with initial momentum $p_0$ and kinematics $z$, $\theta$, and $\phi$, the resulting four-momenta are defined as:
\begin{subequations}
\begin{align}
p_{A} &= p_0\, \{1,\, 0,\, 0,\,1\},\\
p_{B} &= p_0\, 
	\left\{z \sqrt{1 + (1-z)^2 \theta^2},\, -z\,(1 - z)\,\theta\, \cos \phi,\, -z\,(1 - z)\,\theta\, \sin \phi,\, z \right\},
\\
p_{C} &= p_0\,
	\left\{(1 - z) \sqrt{1 + z^2 \theta^2},\, z\,(1 - z)\,\theta\, \cos \phi,\, z\,(1 - z)\,\theta\, \sin \phi,\, 1 - z \right\}.
\end{align}
\end{subequations}
Because the ordering of emissions changes how momentum is conserved, the virtuality ordering is implicitly contained in the expressions for the four-momenta.
While it is possible to express $\Theta_{(1,1)}$ in terms of the splitting kinematics (and we have), it is tedious and unenlightening.

In practice, we use Monte Carlo integration to perform the integral in \Eq{eq:aeascross}.
We generate ``events'' with each parameter $z$ and $\theta$ selected according to a uniform distribution with a lower bound of $0.001$, and $\phi$ distributed uniformly in $[0,2\pi)$.
Each event is assigned a weight equal to the integrand in \Eq{eq:aeascross}.
To implement the plus prescription on $z_g$ in the initial quark case, for each event with an initial quark, we generated a second event with the same values of  $\{z_{\gamma}, \theta_{\gamma}, \theta_g\}$, a negative weight, and $z_g$ selected according to a uniform distribution over $[0,0.001)$.
We use the splitting kinematics to construct three massless four-vectors, after which we use the same \textsc{FastJet} tools as in \Sec{sec:2.4} to implement the isolated photon subjet procedure.

Although the kinematics of \Eq{eq:aeascross} are independent of the jet momentum scale, the parameters $\sigma_q$, $\sigma_g$, and $\alpha_s$ all depend on the momentum.
We performed our analysis at jet transverse momenta of $p_T = \{100, 200, 400, 800\}$ GeV.
The initial quark jet cross section $\sigma_q$ and the gluon jet cross section $\sigma_g$ were determined for each momentum in \textsc{Pythia}. At 400 GeV, we obtained $\sigma_q/\sigma_g = 0.63$.  We assume flavor universality throughout, such that the $\ziso$ distribution does not depend on the quark charges except as a normalization.
At each energy we used a fixed-coupling approximation for the value of $\alpha_s$, evaluated at $\mu = p_T \, R$:
\begin{equation}
	\alpha_s(\mu^2) = \frac{\alpha_s(m_Z^2)}{1 + \alpha_s(m_Z^2) \,\, b_0 \log\left(\frac{\mu^2}{m_Z^2}\right)},
\end{equation}
where $b_0 = (33 - 2 N_f)/(12 \pi)$. Here, $N_f$ is the number of flavors available at the scale $\mu$.

\begin{figure}[t]
    \centering
    \includegraphics{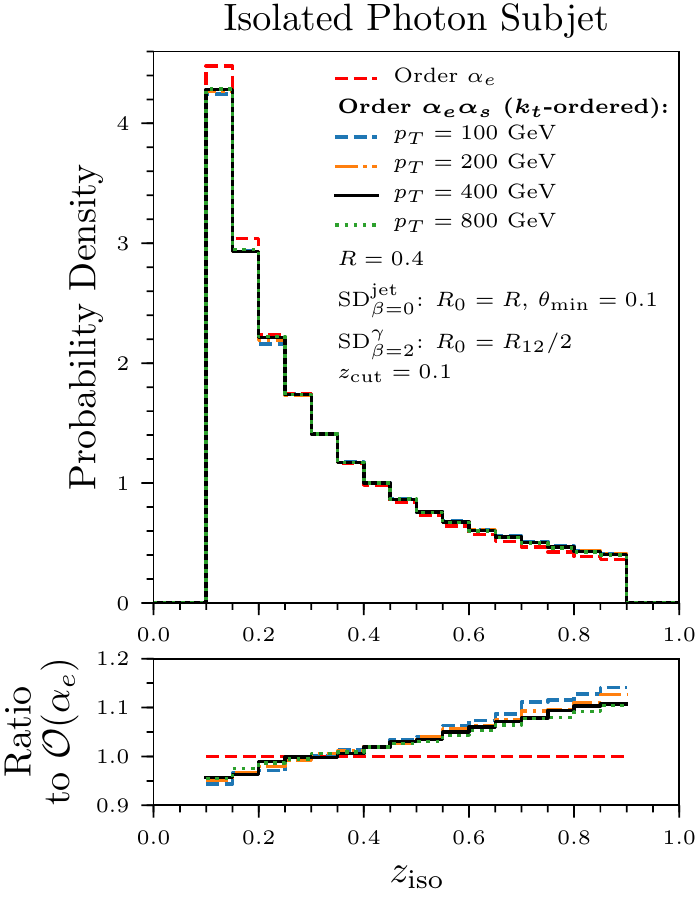}
    \caption{
    Probability densities for isolated photon subjet momentum fraction $\ziso$ at order $\alpha_e$ and order $\alpha_e \alpha_s$ in the collinear limit.
    Shown are results at $p_T = \{100, 200, 400, 800\}$ GeV.}
\label{fig:theor_results}
\end{figure}

In \Fig{fig:theor_results}, we show the order $\alpha_e \alpha_s$ probability densities in $\ziso$.
Compared to the order $\alpha_e$ cross section, the $\alpha_e \alpha_s$ terms yield at most a 10\% suppresion, and as such, the $\ziso$ distribution largely resembles the basic quark-photon splitting function.
The order $\alpha_e \alpha_s$ initial gluon term is for the most part suppressed at a factor of $\sim 0.1$ compared to the order $\alpha_e \alpha_s$ initial quark term and contributes a correction to the order $\alpha_e$ result of at most $1\%$.
Changing the virtuality scale $n$ between $n = 1/2$ and $n = 2$ has an effect of at most $4 \%$, so we expect that including higher-order contributions to the cross section or relaxing the strong-ordering assumption would have a mild impact on the final shape of the distribution.

\subsection{Parton shower study}
\label{sec:3.4}

\begin{figure}[t]
	\centering
	\begin{subfigure}[b]{0.47\textwidth}
	\centering
    \includegraphics{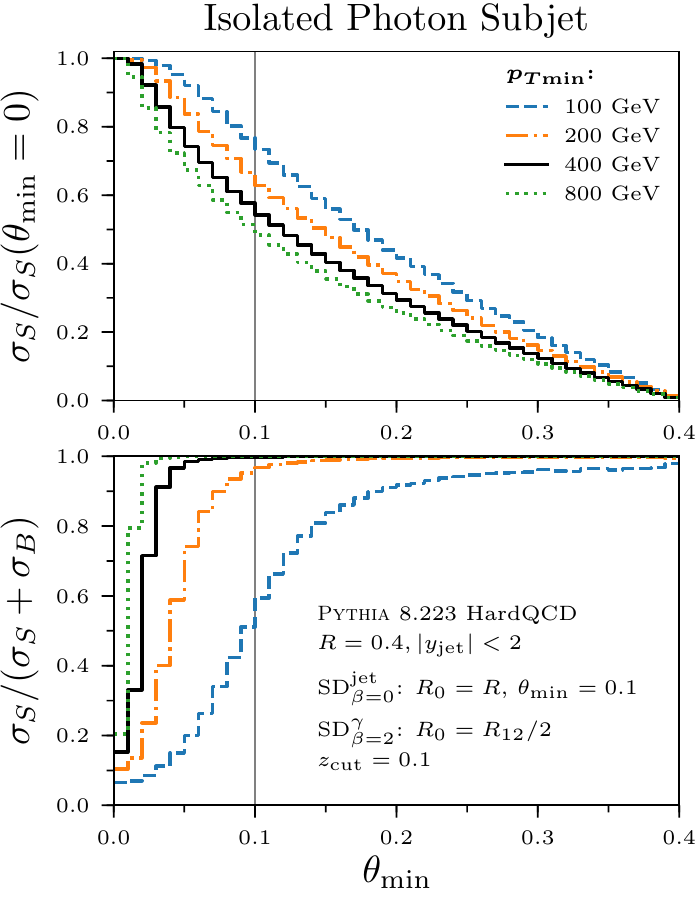}
    \caption{}
    \label{fig:5b}
	\end{subfigure}
	\hfill
	\begin{subfigure}[b]{0.47\textwidth}
	\centering
    \includegraphics{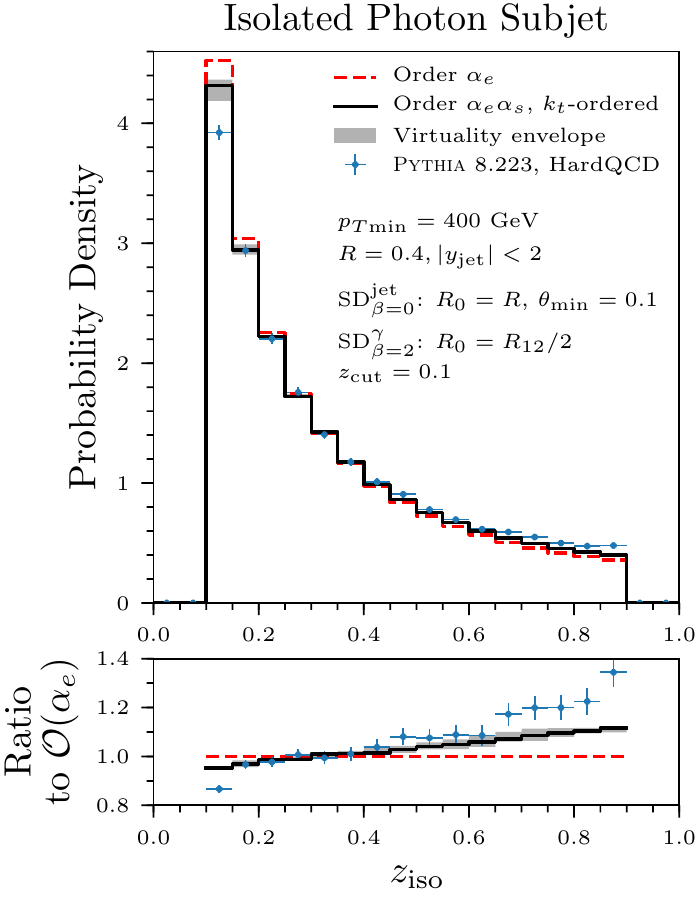}
    \caption{}
    \label{fig:pythia_splitting}
    \end{subfigure}
    \caption{\textbf{(a)} Top: \textsc{Pythia} cross sections of the $q \rightarrow q \gamma$ signal as a function of $\theta_{\rm min}$, given as a ratio to the cross section at $\theta_{\rm min} = 0$. The signal also decreases with $p_T$, and we found $\sigma_S(\theta_{\rm min} = 0) = \{1000, 96, 6.2, 0.22\}$~pb at $p_{T \rm min}= \{100, 200, 400, 800\}$~GeV. This background does not include backgrounds from ``fake'' photons from collinear $\pi^0 \rightarrow \gamma \gamma$ decays. Bottom: ratio of the signal cross section to the sum of signal and background cross sections. \textbf{(b)} Probability distributions of $\ziso$ for the isolated photon subjet at order $\alpha_e$, order $\alpha_e \alpha_s$, and in \textsc{Pythia} with $p_{T\text{min}} = 400$ GeV and $\theta_{\text{min}} = 0.1$. }
\end{figure}

We now perform a parton shower study in \textsc{Pythia} 8.223, with the aim of testing the robustness of the $\ziso$ distribution to hadronization effects.\footnote{At the perturbative level, \textsc{Pythia} has the same formal accuracy as \Sec{sec:3.3} for a single gluon emission in the collinear limit.}
We generate events from the \textsc{HardQCD} process, which encodes $2\rightarrow2$ hard QCD events.
We made event samples for $p_{T\text{min}} = \{100, 200, 400, 800\}$ GeV, each with 20 million events.\footnote{In each case, we set the \textsc{Pythia} parameter $\hat{p}_{T\text{min}}$ to be $20\%$ lower than the jet $p_T$ cut. For the final 400 GeV run in \Fig{fig:pythia_splitting}, we generated 40 million events in order to decrease the statistical uncertainties.}
Because the efficiency for finding isolated photon subjets is so small, we turn off ISR and underlying event to speed up event generation, leaving all other \textsc{Pythia} settings at their default values.
Since the isolated photon subjet condition is based on jet grooming, we do not expect these modifications to make a large impact on our results, though a detailed study of these effects is warranted.  

Events were clustered into anti-$k_T$ jets of radius $R = 0.4$ with a transverse momentum cut $p_{T\text{jet}} > p_{T\text{min}}$ and a rapidity cut $|y_{\text{jet}}| < 2$.
The clustering step and the isolated photon subjet step were implemented using \textsc{FastJet} and \textsc{FastJet Contrib} using the same code for the order $\alpha_e \alpha_s$ calculation in \Sec{sec:3.3}. 
As in \Sec{sec:2.4}, we used \textsc{Pythia} truth information to perform particle identification.

At low energies and low angles, the isolated photon subjet sample was found to be dominated by neutral pion decays: because the observable identifies the photon ``prongs'' of a jet, it was in many cases identifying one of the photons produced in such a decay.
These contributions are relatively easily avoided by choosing appropriate values for $\theta_{\text{min}}$ and $p_{T\text{min}}$; whereas pion decays become more collinear at higher energies, the angular aspect of QED branchings is energy independent.
Using \textsc{Pythia} truth information, we were able to identify signal (photons from QED branchings) and background (all other photons).
In \Fig{fig:5b}, we show signal and background rates for isolated photons at different values of $\theta_{\rm min}$ and $p_{T\text{min}}$.
We choose to use $p_{T\text{min}} = 400$ GeV and $\theta_{\text{min}} = 0.1$ for the remainder of this study, as these values yielded signal cross section of around 3 pb for a background cross section of around $0.006$ pb.
This corresponds to around 150,000 recorded events for the 45 $\text{fb}^{-1}$ 2017 run of CMS \cite{CMSLumi}, of which only about 300 events would be from the pion background.
This value of $\theta_{\text{min}}$ is also a sensible cut from the perspective of the granularity of a typical hadronic calorimeter.

As alluded to in \Secs{sec:2.1}{sec:2.4}, there is also a potential background from closely collinear $\pi^0 \rightarrow \gamma \gamma$ decays, since in a realistic detector it is possible for two nearly-collinear photons to register as a single photon. 
To obtain an approximate sense of this background rate, we relaxed our definition of a photon to include two photons within a distance $\Delta R = 0.025$ from each other, roughly corresponding to the granularity of a typical electromagnetic calorimeter (ECAL).
At 400 GeV with $\theta_{\rm min} = 0.1$, this yielded a background rate of 6\%.
The use of shower-shape observables, which are already well studied at both CMS and ATLAS~\cite{Khachatryan:2015iwa,Aaboud:2016yuq}, would mitigate this background.
To properly quantify this effect, a full study including detector simulation would be necessary.

In \Fig{fig:pythia_splitting}, we show the probability distribution in $\ziso$ for $p_{T\text{min}} = 400$ GeV and $\theta_{\text{min}} = 0.1$ plotted against the corresponding distributions for order $\alpha_e$ and $\alpha_e \alpha_s$ theoretical results.
The \textsc{Pythia} distribution exhibits quite good correspondence with the perturbative results.
It appears that the higher-order corrections are somewhat amplified, albeit with the same functional form.
This is likely due to non-perturbative effects arising from the non-collinear hadronization of the quark subjet, which introduces some soft radiation into the photon subjet.
In order to test the effect of hadronization, we applied the same isolated photon subjet criterion to \textsc{Pythia} events with hadronization disabled and found slightly closer matching to the order $\alpha_e \alpha_s$ distribution.

It is clear from \Fig{fig:pythia_splitting} that, even with higher-order effects, the isolated photon subjet clearly exposes the form of the QED splitting function.
This parton shower study therefore validates the use of isolated photon subjets to expose the splitting function in realistic collider scenarios.

\section{Conclusion}
\label{sec:conclusion}

In the first half of this paper, we introduced soft drop isolation, a new form of photon isolation based on techniques from jet substructure.
Soft drop isolation is infrared and collinear safe and equivalent at leading (non-trivial) order to the most common form of Frixione isolation, making it well suited to perturbative calculations of direct photons.
Soft drop isolation is also democratic and based on clustering algorithms, making it well suited to identify direct photons in jet-rich environments.
Together, these features make soft drop isolation a natural choice for photon studies at the LHC.

In the second half of this paper, we turned to indirect photons, using a combination of soft drop declustering and soft drop isolation to define isolated photon subjets.
We showed how the momentum fraction carried by isolated photon subjets can be used to expose the QED splitting function, which describes the momentum sharing distribution of quark-photon branchings in the collinear limit.
This is a novel test of gauge theories which complements previous soft-drop studies of the QCD splitting function.

As a further extension of this method, soft drop isolation could provide a new way to handle detector granularity.
All collinear-safe isolation criteria are complicated by granularity, which forces the isolation to cut off at the detector's angular resolution when implemented in experiment.
This makes matching between calculations (in which there is no cut-off) and experimental implementations more difficult.
\Ref{Binoth:2010nha} has addresses this issue for Frixione isolation by using a set of concentric cones instead of a smoothly varying cone.
Treating angular resolution with soft drop isolation would be quite straightforward, owing to its clustering basis.
One could introduce a parameter $\theta_{\rm min}$ (analogous to that in \Sec{sec:3}) related to the detector's angular resolution and stop the declustering when the angle between the two subjets was less than $\theta_{\rm min}$.
Because the C/A declustering is angular ordered, this means that the isolation would only treat features with angular separation greater than the detector resolution.
While this is not identical to the behavior in granular detectors, we expect it to closely approximate that behavior.

It is possible to envision a number of extensions to the QED splitting analysis performed in \Sec{sec:3}.
Parallel to the analysis performed in \Ref{Ilten:2017rbd} for the QCD splitting function, the isolated photon subjet criterion could be used in combination with flavor tagging to identify heavy-flavor QED splittings.
Additionally, the same QED splitting analysis could be performed on leptons.
While lepton QED splittings are well studied given the lack of lepton hadronization, it could nevertheless be an interesting test of this new democratic isolation scheme.

Finally, the isolated photon subjet also opens the door to additional photon substructure studies and observables beyond the QED splitting function.
In this paper, we analyzed two-prong substructure with one hadronic subjet and one isolated photon subjet; by recursively applying the soft drop condition \cite{Dreyer:2018tjj}, one could study jets with two (or more) isolated photon subjets.
Such multi-photon configurations could be interesting for studying photon jets \cite{Ellis:2012zp}: jets composed primarily of photons that arise from scenarios beyond the standard model.
Additionally, isolated photon subjets could be used to tag boosted decays such as $h \rightarrow Z \gamma$ or, more broadly, possible decays to jets and photons of boosted beyond-the-standard-model objects.

Isolated photon subjets provide a powerful framework for the study of QED substructure within QCD jets.
We hope that the existence of this technique---and more generally, of a democratic, collinear-safe photon isolation criterion---will encourage the further development of photon-based jet substructure observables.

\begin{acknowledgments}

We thank Fr\'{e}d\'{e}ric Dreyer, Markus Ebert, Stefano Frixione, Andrew Larkoski, Simone Marzani, and Mike Williams for interesting discussions and feedback on this manuscript.
This work was supported by the Office of High Energy Physics of the U.S. Department of Energy (DOE) under grant DE-SC0012567.
The work of ZH was supported by the MIT Undergraduate Research Opportunities Program.

\end{acknowledgments}

\bibliography{softdrop}
\bibliographystyle{JHEP}

\end{document}